\documentclass[%
 preprint,
 amsmath,amssymb,
 aps,
 superscriptaddress,
]{revtex4-1}

\usepackage{graphicx}
\usepackage{dcolumn}
\usepackage{bm}
\usepackage{float}
\usepackage{footnote}
\usepackage{amsmath}
\usepackage[utf8]{inputenc}
\usepackage{array}
\usepackage{makecell}
\usepackage{gensymb}
\usepackage{adjustbox}
\usepackage{ulem}
\usepackage{color}

\begin{document}

\title{Origin of multiple Lifshitz transitions in the Weyl semi-metal RhSi }

\author{Krishnendu Patra}
\affiliation{Department of Condensed Matter and Materials Physics, S. N. Bose National Center for Basic Sciences, Block-JD, Salt lake, Kolkata-700106, India}
\author{Viktor Christiansson}
\affiliation{Department of Physics, University of Fribourg, 1700 Fribourg, Switzerland}
\author{Ferdi Aryasetiawan}
\affiliation{Department of Mathematical Physics, Lund University, Lund, Sweden}
\author{Priya Mahadevan}
\affiliation{Department of Condensed Matter and Materials Physics, S. N. Bose National Center for Basic Sciences, Block-JD, Salt lake, Kolkata-700106, India}


\begin{abstract}
It is known from density functional theory (DFT) calculations that RhSi has a multifold degenerate Dirac point at the Fermi energy, with the dominant states in the low-energy region displaying mostly Rh $d$ character. 
Using DFT+U, we calculate the band structure by considering an effective local interaction on the Rh $d$ states, with a realistic effective Hubbard $U_\textrm{eff}=2.5$ eV derived from a constrained random-phase approximation calculation, and find the emergence of a double hump structure close to the Fermi energy. 
By further deriving a low-energy tight-binding model from our first-principles results, we show that the double hump is a direct consequence of a competition between the Rh $d$-Rh $d$ and Rh $d$-Si $p$ interactions, which differ in their momentum dependence.
As a consequence, through an artificial tuning of the energy level of the Si $p$ orbitals this hump structure can be suppressed due to the effectively reduced Rh $d$ -Si $p$ interaction.
This peculiar low-energy electronic structure additionally results in that a small hole/electron doping ($\sim$ 0.1\%) can tune the Fermi surface topology, going from closed to open Fermi surfaces, which has dramatic consequences for the thermal transport.
 
\end{abstract}

\maketitle

\section{Introduction}

Over the past few years, topological Weyl semimetals have gained significant attention in condensed matter systems. A Weyl semimetal is characterized by a set of points where the bands with opposite curvature touch near the Fermi energy commonly referred to as Weyl points\cite{murakami2007phase,wan2011topological,burkov2011weyl}. These nodes are topologically protected and have a non-zero Chern number and exist in pairs with opposite Chern numbers, which are connected in momentum space by a surface state called the Fermi arc. To generate these Weyl nodes, either time-reversal symmetry or inversion symmetry must be broken. In general, Weyl nodes exhibit a two-fold degeneracy and act as spin-half fermions. They have been identified in various materials, particularly in noncentrosymmetric transition monopnictides like TaAs, NbAs, TaP, NbP\cite{huang2015weyl, PhysRevX.5.011029,xu2015discovery,xu2016observation,lv2015experimental,lv2015observation,yang2015weyl} and magnetic compounds such as Co$_{3}$Sn$_{2}$S$_{2}$ \cite{liu2019magnetic,morali2019fermi} and Co$_{2}$MnGa \cite{belopolski2019discovery}. All of these Weyl semimetals have a small Fermi arc, indicating a narrow non-trivial energy window.However these materials could not be examined for many-body effects associated with the topological boundary states \cite{rao2023charge} because of the coexistence of trivial states in the same energy window.Recently, the RhSi family (RhSi, CoSi, RhGe, and CoGe) with a chiral structure of space group P2$_{1}$3 \cite{geller1954crystal,engstrom1965least} has shown promise. Calculating the electronic structure for these compounds, one finds that Weyl nodes in the bulk exhibit more than a two-fold degeneracy, with the nodes at the zone center ($\Gamma$ point) exhibiting a 3$\times$2 fold degeneracy in the absence of spin-orbit interactions. However, under the influence of spin-orbit interactions, the bands undergo a splitting, resulting in four-fold degenerate chiral fermions at the Fermi level. These can be considered as a combination of two spin-3/2 and two spin-1/2 Weyl fermions. Similarly at the zone boundary (R point), the 4$\times$2 fold degenerate Weyl node behaves as six-fold degenerate double spin-1 fermions in the presence of spin-orbit interactions.\cite{chang2017unconventional,tang2017multiple,huber2022network}. These materials have a wide topologically nontrivial energy window in momentum space, along with long Fermi arcs on their surface \cite{rao2019observation,chang2017unconventional,tang2017multiple,takane2019observation,sanchez2023tunable}. Hence, they are ideal candidates to study the many-body effects of topological boundary states\cite{rao2023charge,rao2019observation}. These characteristics also make them promising candidates for exhibiting circular photogalvanic effect\cite{chang2017unconventional,de2017quantized,rees2020helicity,ni2021giant,chan2017photocurrents,ma2017direct}. Unconventional chiral fermions additionally provide a path for experimental applications, allowing the investigation of materials suited for topological surface states and bulk chiral transport\cite{zhong2016gyrotropic,ma2015chiral,fang2016topological,sanchez2023tunable}. \\
The topological properties of multifold Weyl semimetals have been extensively examined theoretically based on density functional theory \cite{Hohenberg1964,Kohn1965} (DFT) calculation.Recently, STM measurements supported by DFT  calculations have found a unidirectional charge density wave (CDW) on the (001) surface of  CoSi\cite{rao2023charge,li2022chirality}. This CDW, observed in a material that hosts nontrivial topological states is predicted to be driven by electron correlation effects. There have been some studies using DFT + dynamical mean-field theory (DFT+DMFT) for CoSi that find a modification of the band dispersion with a coexistence of coherent and incoherent features \cite{Dutta2019,dutta2021electronic}. However, there is largely a lack of studies in the literature making a direct connection between the unconventional chiral fermions and electronic correlations effects.\\
In this paper, we have chosen RhSi as a representative example of the family and explore the impact of incorporating an on-site (local) effective Coulomb repulsion in DFT+U calculations. For this purpose, we have used the constrained random phase approximation (cRPA) method \cite{aryasetiawan2004frequency} to determine a realistic value for the effective bare interaction U for the Rh $d$ states, which are the main contributors to the low-energy window from 6 eV below the Fermi energy to around 3.5 eV above. Our calculations show an unusual electronic structure, with a double hump feature emerging in the band structure close to the Fermi energy. This opens up for the possibility of multiple consecutive Lifshitz transitions close to the Weyl point at $\Gamma$ through a slight tuning of the chemical potential. Since a Lifshitz transition is an electronic transition that occurs due to a change in the Fermi surface topology with a continuous change in some external parameter\cite{lifshitz1960anomalies}, we directly explore its effect through an effective hole and electron doping (by a rigid shift of the chemical potential). Since such a transition is known to induce abrupt changes in some thermodynamic coefficients\cite{vaks1981singularities,lifshitz1960anomalies} we have additionally calculated the Seebeck coefficient to probe the transition.  
Additionally, by further studying a realistic tight-binding model, we identified the origin of this Lifshitz transition at a microscopic level as a competition between Rh $d$-Rh $d$ and Rh $d$-Si $p$ interactions.
The transition can be of further importance, especially for Weyl points\cite{yang2019topological,xu2018evidence}, as it is sometimes accompanied by various correlated electronic phases that it could be associated with, e.g., a Van Hove singularity which has been related to superconductivity \cite{norman2010lifshitz,shi2017enhanced}. 


\section{Methodology}
We have used the experimental simple cubic structure of RhSi with space group P2$_{1}$3 \cite{geller1954crystal} for our study. The electronic structure was calculated using the DFT+U method with the plane-wave basis set implementation within the Vienna ab initio simulation package (VASP), using projected augmented wave potentials \cite{blochl1994projector} \cite{kresse1993ab,kresse1994ab,kresse1996efficient}.
The generalized gradient approximation (GGA) was used for the exchange-correlation functional\cite{perdew1996generalized}, with onsite Columb repulsions U$_{eff}$ applied on the Rh $d$ orbitals using the implementation by Dudarev \cite{dudarev1998electron}. The appropriate U on the Rh $d$ states was determined from a cRPA \cite{aryasetiawan2004frequency} calculation. Here, we remove only the screening channels coming from the $d$ states in the polarization function $\Pi^\textrm{r}=\Pi-\Pi^d$ and calculate an effective bare interaction as $U(\omega)=[1-v\Pi(\omega)]^{-1}v$. While in principle frequency dependent, we take the static value as our Hubbard $U=3.1$ eV and Hund's coupling $J=0.6$ eV for the DFT+U calculation, giving an effective $U_\textrm{eff}=U-J=2.5$ eV. We have optimized all the internal atomic positions through a total energy minimization scheme, while keeping the lattice constants fixed at the experimental value. A cutoff energy of 400 eV was considered to determine the maximum kinetic energy cut-off for the plane waves included in the basis. Integrations over the Brillouin zone were carried out using a Monkhorst Pack\cite{monkhorst1976special} $k$-grid of $9\times 9 \times 9$. 

To further investigate the contribution of different interactions to the low-energy electronic structure, we mapped our DFT band structure onto a tight binding model using a basis set of maximally localized Wannier functions, with Rh $d$- and Si $p$-like orbitals included in the basis. The matrix elements were calculated using the wannier90-VASP interface \cite{mostofi2008wannier90,franchini2012maximally}.  As we have a reasonably good fit between our first principles band structure and model one, with the orbital basis well represented by Wannier functions, we use this tight-binding Hamiltonian for our further analysis.

 By varying the chemical potential (mimicking electron or hole doping), we study how the Seebeck coefficient is affected to monitor the resulting changes in thermodynamic properties. 
This was calculated within the semiclassical Boltzmann transport equation by interpolating the DFT band dispersions. The calculations were done using the relaxation time approximation as well as a rigid band approximation, as implemented in Boltztrap2 \cite{madsen2018boltztrap2}. 
The corresponding formula utilized  to compute Seebeck coefficients at an absolute temperature T is given by\cite{madsen2006boltztrap,gupta2019theoretical},
\begin{equation}
\text{S} = \frac{e}{\text{T} \sigma}\int^{-\infty}_{\infty}\text{F}(\text{E,T})(\text{E}-\mu)(\frac{\partial \text{f}_{0}}{\partial \text{E}})\text{dE}
\end{equation}
Here $\mu$ is the chemical potential,$f_0$ is the Fermi function, $e$ is the electron charge, and $\sigma$ is the electrical conductivity, defined as
\begin{equation}
\sigma = e^{2}\int^{\infty}_{-\infty}\text{F(E,T)}(-\frac{\partial \text{f}_{0}}{\partial \text{E}}) \text{dE} 
\end{equation}
The transport distribution function is calculated as $\text{F(E,T)} = \int \text{v}_{\text{k}}\otimes \text{v}_{\text{k}} \tau_{\text{k}} \delta (\text{E}-\text{E}_{\text{k}}) \frac{\text{dk}}{8\pi^{3}} $. Here, $\text{v}_{\text{k}}$ is the component of group velocity in a particular direction, $\text{E}_{\text{k}}$ is the Kohn-Sham energy eigenvalues of state $k$, and $\tau_{\text{k}}$ is the total relaxation time.

\section{Result and Discussion}
\begin{figure}[hbt]
\centering
\includegraphics[width=1.05\columnwidth]{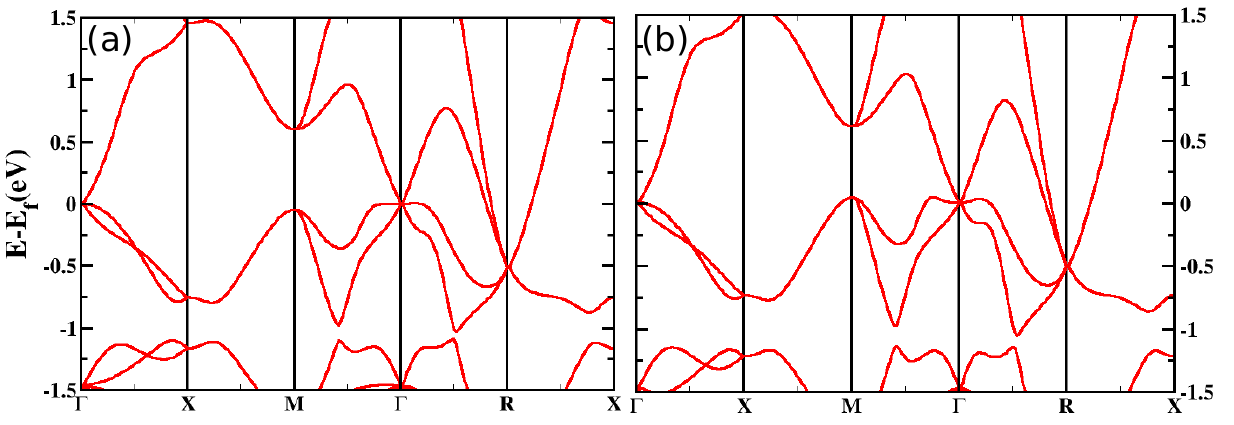}
 \caption{The calculated electronic band dispersions within GGA+U for (a) $U= 0.0$~eV and (b) $U= 2.5$~eV. E$_{f}$ represents the Fermi energy.}
\label{fig1}
\end{figure}

The low-energy electronic structure of RhSi, calculated within GGA+U, is shown in Figure Fig.~\ref{fig1}. In  Fig.~\ref{fig1}(a), we show the calculated band dispersion for $U=0$ along the high-symmetry directions. It contains a multifold degenerate Dirac point, which has a degeneracy of eight at the R point and six at the zone center (in the presence of spin-orbit interactions, these are reduced to six at the R point and four at $\Gamma$).  We also find a flat band along the M-$\Gamma$-R path, in the vicinity of the Fermi energy.

These findings are consistent with previous studies \cite{chang2017unconventional}. As strong correlation effects in other members of this family of materials have been found to modify the DFT picture \cite{Dutta2019,dutta2021electronic} and also lead to charge density waves being seen on the surface of CoSi\cite{rao2023charge,li2022chirality}, we go on to determine the appropriate value of $U$ for the system by using cRPA. These calculations suggest an appropriate value of $U_\textrm{eff}=U-J=2.5$ eV. Using this value in our GGA+U calculations, we obtain the corresponding band dispersion plotted in panel b of Fig.~\ref{fig1}. we find that incorporating a finite $U$ preserves both Weyl points in the electronic structure without disrupting any crystal symmetry.  However, there is a notable change in the band dispersion in its vicinity near the $\Gamma$ point, visible along the M-$\Gamma$-R path. Instead of the nearly flat band observed for $U=0$, a double-hump pattern emerges in the narrow energy window close to the Fermi energy. The presence of this double-hump structure suggests the occurrence of a Lifshitz transition\cite{ito2019enhanced,okamoto2010discontinuous}, which can be induced by a tuning of certain parameters (e.g., pressure or doping).

\begin{figure}[hbt]
\centering
\includegraphics[width=.95\columnwidth]{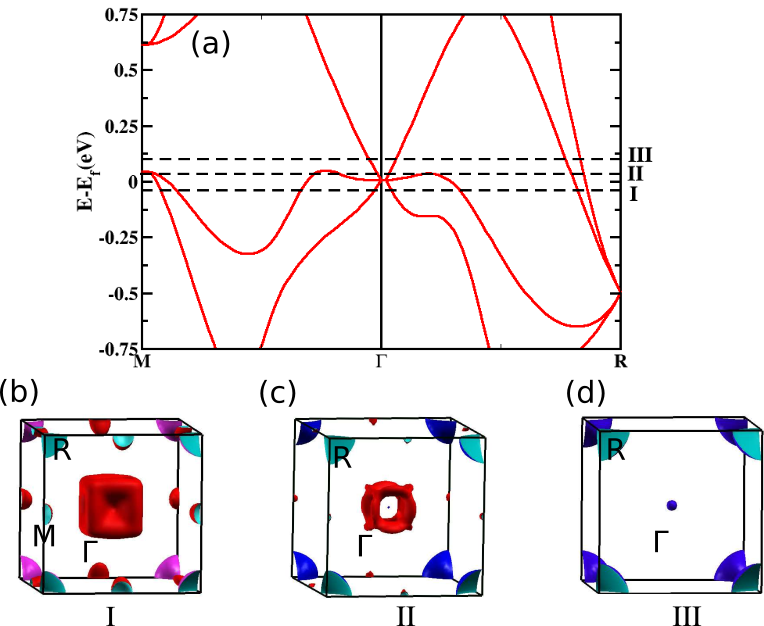}
 \caption{(a) Electronic band dispersions along the M-$\Gamma$-R path for the different Fermi energy labeled I, II and III. These correspond to  0.1$\%$ hole doping, 0.1$\%$ electron doping, and 0.2$\%$ electron doping, respectively. The corresponding Fermi surfaces are shown in panels (b), (c) and (d) respectively.}
 \label{fig2}
\end{figure}

In the present case, the transition is driven by tuning the chemical potential through carrier doping. Considering a rigid band approximation, three different locations for the Fermi level (labeled I, II, and III) associated with three different percentages of carrier doping are shown in Fig.~\ref{fig2}(a). Label I refers to 0.1 $\%$ hole doping while labels II and III correspond to 0.1$\%$ and 0.2$\%$ electron doping, respectively. Fermi levels I and II enclose the Weyl point at $\Gamma$ and Fermi levels II and III almost enclose the maxima points of the double hump.These points are expected to be the transition points. Fig.~\ref{fig2}(b-d) display the Fermi surfaces corresponding to the three different doping concentrations. The evolution of the Fermi surfaces between the three distinct energy levels is clearly visible.At label I, we have a hole pocket(red) which encompasses a closed region around the $\Gamma$ point whereas at label II we have an open hole pocket with an additional electron(blue) pocket emerging inside it.  At label III, the hole Fermi pocket disappears, and we have an enlarged electron pocket. These changes in the topology of the Fermi surface at the three distinct labels signify multiple successive Lifshitz transitions. 

\begin{figure}[h]
\centering
\includegraphics[width=.95\columnwidth]{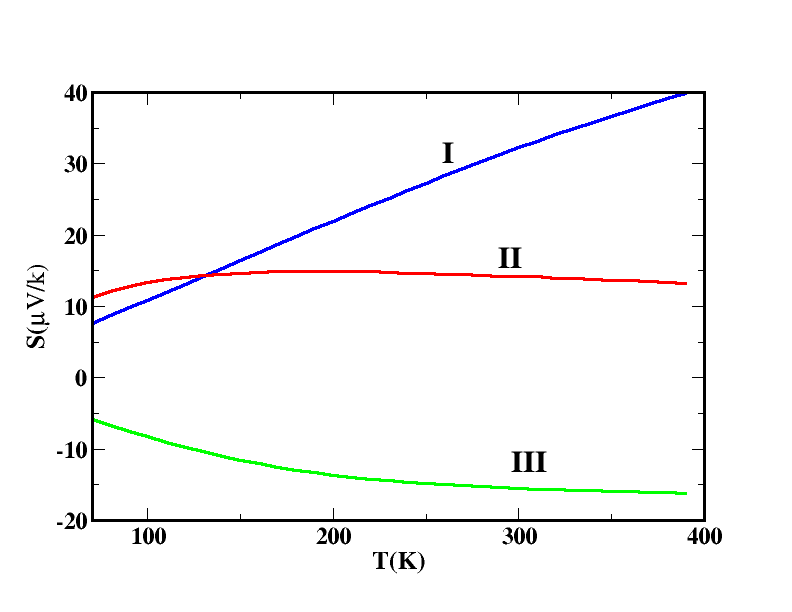}
 \caption{The Seebeck Coefficient as a function of temperature for the chemical potentials labeled  I, II, and III in Fig.~\ref{fig2}.These correspond to the doping percentages  0.1$\%$ hole doping (blue line), 0.1$\%$ electron doping (red) and 0.2$\%$ electron doping (green), respectively.}
\label{fig3}
\end{figure}

Various thermodynamic measurements can be used to probe Lifshitz transitions. The Seebeck coefficient is one such parameter that depends on the Fermi surface topology. In Fig.~\ref{fig3} we have plotted the Seebeck coefficient as a function of temperature with the different doping percentages shown previously in Fig.~\ref{fig2}. At I ($0.1\%$ hole doping), the Seebeck coefficient is positive and roughly linear within the specified temperature window, indicating a majority of carriers as holes. In Region II ($0.1\%$ electron doping), although the Seebeck coefficient is still positive, it exhibits a different trend. Initially, it rises with temperature and then nearly flattens out. At III ($0.2\%$ electron doping), the Seebeck coefficient shows negative values, which indicates that a majority of the carriers are now electrons. It initially decreases with temperature and then again becomes roughly flat. Hence we find dramatic changes in the Seebeck coefficient with only small changes in the doped carrier concentration as a direct result of the peculiar changes to the low-energy electronic structure.

\begin{figure}[h]
\centering
 \includegraphics[width=.75\columnwidth]{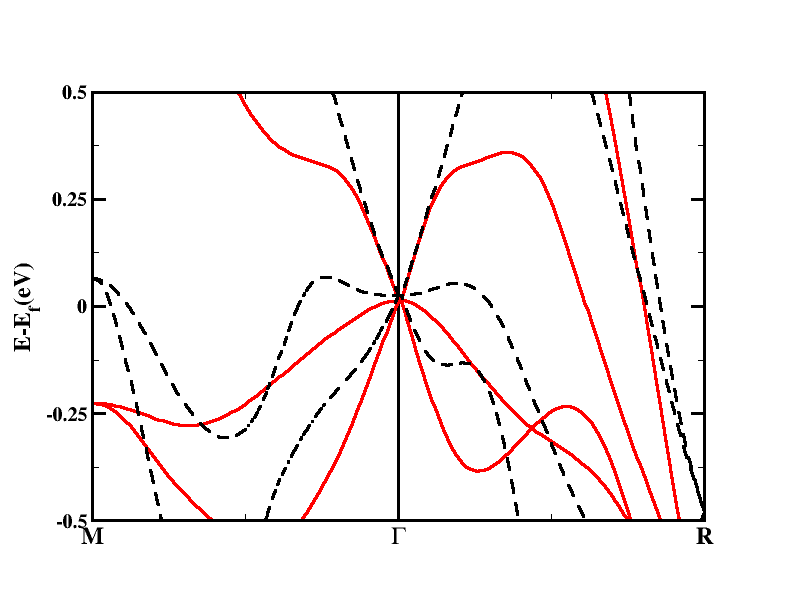}
\caption{Electronic band dispersion along M-$\Gamma$-R path for the tight-binding model of RhSi with the onsite energies of Si-p orbitals decreased by 10 eV(red solid line) compared with the original band structure(black dashed line).The Weyl points at $\Gamma$ have been aligned for both plots.}
\label{fig4}
\end{figure}

To further unravel the origin of the double hump pattern that we find in the band structure, which 
as discussed above gives rise to multiple Lifshitz transitions, we mapped the ab initio band structure onto a tight-binding model that includes Rh $d$ and Si $p$ states in the basis. 
We have obtained a good fit of the \textit{ab-initio} band structure, and the comparison as well 
as the spread of each Wannier function are given in the Supplementary Information.
Utilizing this tight-binding model, we can artificially tune various parameters to explore different interaction pathways. In this scenario, we observed that by systematically reducing the on-site energies of the Si-$p$ orbitals, we can modify the double hump pattern. The resulting band dispersion, obtained by reducing the on-site energies of Si-$p$ orbitals by 10~eV, are shown in Fig.~\ref{fig4}. We find that the double hump pattern near $\Gamma$ point almost vanishes, and furthermore does not regain its flat-band-like appearance.  
Whereas just an examination of the \textit{ab-initio} band dispersion in terms of the character of states would reveal that the contribution of Rh-$d$ states dominates over Si-$p$ states near the Fermi energy, and could be expected to dominate the physics in this energy window. Our results therefore show that the Rh $d$-Si $p$ interactions further perturb the electronic structure. The competing contributions from Rh $d$-Rh $d$ and Rh $d$-Si $p$ orbitals, which have a different momentum $k$ dependence, therefore give rise to the double hump pattern. Artificially reducing the on-site energies of the Si-$p$ orbitals effectively pushes the Si states deeper into the valence band, further away from the Fermi energy. Consequently, the effective Rh $d$-Si $p$ interaction strength decreases, explaining why the double hump pattern vanishes.

\section{Conclusions}

Since the nonlocal interaction between the Rh $d$ and Si $p$ states was found to play a role in the low-energy description, it would be of interest to include also an effective nonlocal interaction. This could be achieved for example in DFT+U+V, or with a more numerically advanced scheme such as $GW$+EDMFT \cite{Biermann2003} treating both local and nonlocal interaction and screening effects.

\section{Acknowledgments}

The authors acknowledge the assistance received from DST through the project no DST/INT/SWD/VR-P-08/2019. The 
support provided by PARAM Shivay Facility under the National Supercomputing Mission, Government of India at the Indian Institute of Technology, Varanasi is highly acknowledged.

\bibliography{reference.bib}

 \end{document}